# Tables of Einstein coefficients and lifetimes of upper rovibronic levels for Q-branch lines of the $d^3\Pi_u^-, v' \to a^3\Sigma_g^+, v''$ bands for the H₂, HD, D₂, and T₂ molecules.

© 2015г. **B. P. Lavrov, L. L. Pozdeev, V. I. Yakovleva**

*Sankt-Petersburg State University*

*E-mail: b.lavrov@spbu.ru*

## Abstract

The present work reports results of the semi-empirical determination of the spontaneous emission transition probabilities (Einstein coefficients) for Q-branch lines of Fulcher-α band system and radiative lifetimes of upper electronic-vibro-rotational (rovibronic) levels for the most important isotopologues of diatomic hydrogen. They are based on an adiabatic theoretical model and all currently available experimental data on rovibronic energy levels, ratios of the line strengths and the lifetimes. Numerical data are presented in tabular format for vibrational quantum numbers $v'$=0-6 and $v''$=0-7 ($N$ is the quantum number of total angular momentum of the molecule excluding electron and nuclear spins). The uncertainties (one SD) of experimental and semi-empirical data are listed for each datum. Currently available results of *ab initio* calculations are listed and may be used for comparisons. For the $d^3\Pi_u^-, v'$=4÷6,$N$=1 levels of the H₂ molecule the lifetimes caused by predissociation via non-adiabatic interaction with continuum of the $c^3\Pi_u^-$ electronic state were estimated for the first time in the basis of experimental and semiempirical data.

**Key words:** hydrogen; deuterium; tritium; excited electronic states; rovibronic levels; transition probabilities; Einstein coefficients; mean lifetime; radiative lifetime; non-empirical calculations; predissociation.

**CONTENTS**

# NOMENCLATURE

$A_{av''N}^{dv'N}$ – The spontaneous emission transition probability (Einstein coefficient A) for the Q-branch lines of the ($v'-v''$) band of the $d^3\Pi_u^- \to a^3\Sigma_g^+$ electronic transition.

$v$ – Vibrational quantum number.

$N$ – Quantum number of total angular momentum of the molecule excluding electron and nuclear spins.

$v_{av''N}^{dv'N}$ – The wavenumber of the Q$N$ ($N$= 1, 2, 3…) spectral line in the ($v'-v''$) band of the $d^3\Pi_u^- \to a^3\Sigma_g^+$ Electronic transition.

$r$ – The internuclear separation.

$S_{av''}^{dv'}/S_{av'''}^{dv'}$ – Branching ratios for the Q-branch lines of $v''$–progressions of Fulcher-α bands

$\tau_{dvN}$ – Radiative lifetimes of the $d^3\Pi_u^-, v, N$ rovibronic levels caused by spontaneous decay.

PS – Phase Shift method
MH – Method Hanle
OMR – Molecular Optical Magnetic Resonance
DC – Method of Delayed Coincidences
plasma – Plasma experiments
AA – Calculations based on Adiabatic Approximation
NAM – Calculations based on Non-Adiabatic Model
Expt – Average of most reliable experimental data
SDRL – Semiempirical Determination of Radiative Lifetimes
Diss – Estimate of the lifetime caused by predissociation

## List of Tables

1. Optimal values of parameters for Hulbert-Hirschfelder potential curves of the $a^3\Sigma_g^+$ and $d^3\Pi_u^-$ electronic states of hydrogen molecule obtained in [14]. Standard deviations are shown in brackets in units of least significant digit.

2. Branching ratios $S_{av''}^{dv'}/S_{av'''}^{dv'}$ for Q-branch lines of Fulcher-α bands of $H_2$ and $D_2$ reported in various papers. Asterisks mark experimental data recognized as outliers during semiempirical analysis. Standard deviations are shown in brackets in units of least significant digit.

3. The lifetimes $\tau_{dvN}$ of the $d^3\Pi_u^-, v, N$ electronic-vibro-rotational levels of the $H_2$ molecule reported in various papers. Most reliable experimental data are in bold face; their averages (Expt) are italicized.

4. The lifetimes $\tau_{dvN}$ of the $d^3\Pi_u^-, v, N$ electronic-vibro-rotational levels of the $D_2$, HD and $T_2$ molecules reported in various papers.

5. Einstein coefficients for the $d^3\Pi_u^-, v', N=1 \to a^3\Sigma_g^+, v'', N=1$ electronic-vibro-rotational spontaneous transitions of the $H_2$ molecule, obtained semiempirically in the present work and results of non-empirical adiabatic calculations (in italics) reported in [21]. The uncertainties of the semiempirical determination (one standard deviation) are shown in brackets in units of least significant digit.

6. Einstein coefficients for the $d^3\Pi_u^-, v', N=1 \to a^3\Sigma_g^+, v'', N=1$ electronic-vibro-rotational spontaneous transitions of the HD molecule, obtained semiempirically in the present work and results of non-empirical adiabatic calculations (in italics) reported in [21]. The uncertainties of the semiempirical determination (one standard deviation) are shown in brackets in units of least significant digit.

7. Einstein coefficients for the $d^3\Pi_u^-, v', N=1 \rightarrow a^3\Sigma_g^+, v'', N=1$ electronic-vibro-rotational spontaneous transitions of the $D_2$ molecule, obtained semiempirically in the present work and results of non-empirical adiabatic calculations (in italics) reported in [21]. The uncertainties of the semiempirical determination (one standard deviation) are shown in brackets in units of least significant digit.

8. Einstein coefficients for the $d^3\Pi_u^-, v', N=1 \rightarrow a^3\Sigma_g^+, v'', N=1$ electronic-vibro-rotational spontaneous transitions of the $T_2$ molecule, obtained semiempirically in the present work and results of non-empirical adiabatic calculations (in italics) reported in [21]. The uncertainties of the semiempirical determination (one standard deviation) are shown in brackets in units of least significant digit.

# 1. Introduction

The present work is devoted to semiempirical studies of electronic-vibro-rotational (rovibronic) spontaneous emission transition probabilities for the Q-branch lines of Fulcher-α band system (the $d^3\Pi_u^-, v', N \rightarrow a^3\Sigma_g^+, v'', N$ rovibronic transitions) and radiative lifetimes of upper rovibronic levels for the most important isotopologues of diatomic hydrogen. This band system takes a special place among other band systems located in visible part of a spectrum. On the one hand, it is the most studied band system, already enabled to reveal many effects of intra-molecular dynamics (Λ-doubling, electronic-rotational interaction and predissociation of some part of rovibronic levels). However, it, still, is of interest for quantum chemistry because an accuracy of modern *ab initio* calculations of rovibronic energy levels of the simplest diatomic molecule is significantly less than precision of experimental data. On the other hand, it is the brightest and easily recognizable band system in visible spectrum of hydrogen isotopologues. Therefore spectral lines of Q-branches were proposed to be used for spectroscopic determination of the translational (gas) [1,2] and vibrational [3,4] temperatures, a dissociation degree [5] and isotope content [6] of molecules in non-equilibrium plasmas, containing isotopologues of diatomic hydrogen. Currently these lines are most widely used in spectroscopic diagnostics of ionized gases and plasmas [7].

The goal of the present work was the semi-empirical determination of Einstein coefficients for the $d^3\Pi_u^-, v', N \rightarrow a^3\Sigma_g^+, v'', N$ spontaneous transitions and the lifetimes of upper rovibronic levels of the $H_2$, HD, $D_2$, and $T_2$ molecules in the framework of an adiabatic approximation. The results thus obtained are based on all currently available experimental data about rovibronic term values, ratios of the transition probabilities in the v''-progressions of bands, and lifetimes of the $d^3\Pi_u^-, v', N$ rovibronic levels for the $H_2$ and $D_2$ molecules.

## 2. Semiempirical determination of rovibronic transition probabilities and the lifetimes.

For the determination we use the approach first proposed and realized in [8], which was later developed by including: an optimization procedure of getting r-dependence of the dipole moment of electronic transition [9-10]; the semi-empirical determination of adiabatic potential curves [11]; and accurate calculation of the vibronic matrix elements [12,13].

In an adiabatic approximation Einstein coefficients for the Q-branch lines in Fulcher-α bands may be presented in the form of

$$A_{av"N}^{dv'N} = \frac{32\pi^4}{3h} \left(v_{av"N}^{dv'N}\right)^3 S_{a'v"}^{dv'}, \tag{1}$$

where wavenumbers of lines are connected with energy differences between combining rovibronic states (in cm-1) by well-known relation

$$v_{av"N}^{dv'N} = E_{dv'N} - E_{av"N},$$

and the line strengths of rovibronic transitions are practically independent on N (see [8,12]

$$S_{av"N}^{dv'N} = \left|\langle a,v",N|M_{da}(r)|d,v',N\rangle\right|^2 \approx S_{av"}^{dv'}. \tag{2}$$

Here $M_{da}(r)$ is the so-called electronic transition moment – matrix element of the dipole moment operator on electronic wave functions of the $1s\sigma 3p\pi\ ^3\Pi_u^-$ and $1s\sigma 2s\sigma\ ^3\Sigma_g^+$ adiabatic electronic states; and $|d,v',N\rangle$, $\langle a,v",N|$ - vibrational wave functions of upper and lower rovibronic states.

In contrast with well-known *r*-centroid method, we approximated the $M_{da}(r)$ function by following power series expansion

$$M_{da}(r) = M_0(1 + \sum_{k=1}^{m} a_k(r-r_0)^k), \tag{3}$$

where $M_0$ and $a_1, a_2…, a_m$ are adjustable parameters of the $M_{da}(r)$ approximation in the neighborhood of the $r_0$ point.

Thus, we used following expressions for Einstein coefficients of the $d^-,v',N \rightarrow a,v",N$ rovibronic transitions

$$A_{av"N}^{dv'N} = \frac{32\pi^4}{3h} \left(v_{av"N}^{dv'N}\right)^3 M_0^2 q_{av"}^{dv'}\left[1 + \sum_{k=1}^{m} a_k M_k(v',v",r_0)\right]^2 \tag{4}$$

and for the radiative lifetimes of the $d^3\Pi_u^-, v', N$ rovibronic levels

$$\tau_{dv'N} = \left(\sum_{v''} A_{av''N}^{dv'N}\right)^{-1}, \tag{5}$$

where

$$q_{av''}^{dv'} = |\langle a, v'', N | d, v', N \rangle|^2$$

are Frank-Condon factors and

$$M_k(v', v'', r_0) = \frac{\langle a, v'', N | (r - r_0)^k | d, v', N \rangle}{\langle a, v'', N | d, v', N \rangle} \tag{6}$$

corresponding ratios of matrix elements on vibrational wave functions of combining rovibronic states. Vibrational wave functions (depending on $N$ via centrifugal term in Hamiltonian) needed for calculating these matrix elements were obtained by numerical solution of the vibrational Schrödinger equation. For that we used Hulbert-Hirschfelder potential curves obtained in [14] by the optimization technique [11] from experimental rovibronic term values of the $H_2$ [15] and $D_2$ [16] molecules. Parameters of the Hulbert-Hirschfelder potential reported in [14] are shown in Table 1. The computer code presented in [13] was use for calculating the ratios (6).

From formulae (1-2) one may see that in our theoretical model (based on experimental results reported in [8, 12] the N-dependences of the transition probabilities for Q-branch lines is mainly caused by cubes of their wavenumbers. Therefore, in optically thin plasma the ratios of the line strengths (often called *the branching ratios*) for pairs of lines having the same upper rovibronic level but belonging to two different bands of the v''-progression are connected with corresponding line intensities and wavenumbers in the following way:

$$\frac{S_{av''}^{dv'}}{S_{av'''}^{dv'}} = \left(\frac{v_{av'''N}^{dv'N}}{v_{av''N}^{dv'N}}\right)^3 \frac{I_{av''N}^{dv'N}}{I_{av'''N}^{dv'N}}, \tag{7}$$

where the line intensities $I_{av''N}^{dv'N}, I_{av'''N}^{dv'N}$ are expressed in number of photons emitted by a unite volume per unit time in all directions.

In the present work for measurements of the Q-branch line intensities, we used experimental setup analogous to that described in our previous paper [17]. Plasma of capillary-arc discharge lamps filled with hydrogen (DVS-25 [18]) and deuterium (LD2-D [19]) was used as a light source. For determination of the branching ratios (7), we measured line profiles of the first five Q-branch lines ($N$=1-5) for every pair of the (v′–v″) and (v′–v‴) bands. Then we calculated the line intensities by numerical integration, analyzed the intensity distributions in both branches and putted away outliers caused by overlapping of the lines of interest with neighboring lines belonging to other band systems.

New set of measurements confirmed our previous results reported in [8,12] that the possible dependence of experimental ratios (7) on the quantum number $N$ is within error bars and could be neglected. The data thus obtained are presented in Table 2 together with experimental and semiempirical results reported in [12]. One may see that the difference between new (more precise and reliable) and old [12] data in most cases in within error bars. The branching ratios calculated in the framework of non-adiabatic model [20], as well as those obtained from results of adiabatic calculations reported in [21] are included into Table 2.

One may see fairly good agreement between experimental and calculated values. It should be stressed, that the calculations of Einstein coefficients and radiative lifetimes of electronic-vibrational levels of isotopologues of diatomic hydrogen reported in [21] are based on currently most precise *ab initio* calculations of potential curves of the $1s\sigma 3p\pi\ ^3\Pi_u^-$, $1s\sigma 2s\sigma\ ^3\Sigma_g^+$ adiabatic electronic states and the $r$-dependence of the electronic transition moment $M_{da}(r)$ reported in [22]. Authors of [21] made their calculations for the nonrotating-molecule; i.e. they omitted the centrifugal term in Hamiltonian of the vibrational Schrödinger equation, what leads to about 10 cm$^{-1}$ decrease for the value of lowest vibro-rotation (v=0, $N$=1) energy level of the $1s\sigma 3p\pi ^3\Pi_u^-$ adiabatic electronic state. Nevertheless, the branching ratios (7) calculated in the present work by means of formulae (1), (2) with their data on electronic-vibrational energy levels and radiative transition probabilities $A_{av''N}^{dv'N}$ are in reasonable accordance with our experimental data.

All known to authors of the present work experimental, semiempirical and non-empiric data for lifetimes $\tau_{dvN}$ of the $d^3\Pi_u^-$,v,$N$ electronic-vibro-rotational levels of the $H_2$, HD, $D_2$, and $T_2$ molecules reported in various papers [8-9,20-21,23-33] are listed in Tables 3,4. One may see that currently the experimental data are available only for the $H_2$ and $D_2$ isotopologues. Moreover, for the $D_2$ molecule experimental data are reported by only one research group [20, 33] and have no independent confirmation. Therefore, in the present work they are not used for semiempirical determination of the absolute value of the transition moment $M_{da}(r)$. Experimental data for the $H_2$ isotopologue were examined in the framework of an analysis similar to that recently reported in [34]. Most reliable (from our point of view) experimental data are printed in bold face; their averages (Expt) are italicized.

The effect of qualitatively different mechanisms of depopulating the $d^3\Pi_u^-, v, N$ rovibronic levels of the H$_2$ molecule in low temperature plasma was observed in [35]. The levels with $v$=0–3 located below energy of dissociation of the molecule into two atoms in states with principle quantum numbers n=1 and 2 are depopulated due to spontaneous emission and quenching in collisions with H2 molecules (the rate proportional to their density). The levels with $v$ =4–6, which are located higher than that dissociation limit, show sufficiently higher depopulation rate independent on the gas pressure. In work [31] the assumption is put forward, that this effect is caused by non-adiabatic interaction between the $(1s\sigma 3p\pi)$ $d^3\Pi_u^-$, $v$=4-6, $N$ levels and the continuum of the $(1s\sigma 2p\pi)$ $c^3\Pi_u^-$ electronic state. Based on experimental data reported in [35] and the assumption of existence of that intramolecular process the lifetime of the $d^3\Pi_u^-$, $v$=4-6 levels caused by predissociation was estimated as (16±2) ns. This hypothesis and the estimate later have been confirmed by results the gas-beam experiments [32, 33] and non-adiabatic calculations [20]. At the same time, experimental data and results of calculations reported in [20] show weakness of this effect for the D$_2$ molecule (see Table 4).

Thus for semiempirical determination of absolute values of Einstein coefficients for the $d^3\Pi_u^-, v', N \to a^3\Sigma_g^+, v'', N$ spontaneous transitions and the lifetimes of upper rovibronic levels of the H$_2$, HD, D$_2$, and T$_2$ molecules we used the averaged values of most reliable experimental lifetimes of the $d^3\Pi_u^-, v$=0-3, $N$ levels of the H$_2$ molecule.

Optimum set of adjustable parameters $M_0$ and $a_1, a_2\ldots, a_m$ of the $M_{da}(r)$ approximation and a corresponding variance-covariance matrix are found by minimization of the weighed root-mean-square deviation between experimental $Y_i^{expt}$ and calculated $Y_i^{calc}$ values of branching ratios (7) and radiative lifetimes

$$\Phi(M_0, a_1, \ldots, a_m) = \frac{1}{L_{max} - (m+1) - 1} \sum_i^{L_{max}} \left( \frac{Y_i^{expt} - Y_i^{calc}}{\sigma_i^{expt}} \right)^2, \qquad (8)$$

where $L_{max}$ is total number of active experimental data. If $L$ denotes the number of measured branching ratios, the number of active experimental lifetime values is $(L_{max} - L)$. Thus for $1 \leq i \leq L$

$$Y_i^{expt} = \left( \frac{v_{av'''N}^{dv'N}}{v_{av''N}^{dv'N}} \right)^3 \frac{I_{av''N}^{dv'N}}{I_{av'''N}^{dv'N}}, \qquad (9)$$

$$Y_i^{calc}(r_0, a_1, ..., a_m) = \left( \frac{q_{av''}^{dv'}\left[1 + \sum_{k=1}^{m} a_k M_k(v', v'', r_0)\right]^2}{q_{av'''}^{dv'}\left[1 + \sum_{k=1}^{m} a_k M_k(v', v''', r_0)\right]^2} \right) . \qquad (10).$$

For $L < i \leq L_{max}$ lifetime values from the row *Expt* of Table 3 are used as $Y_i^{expt}$, and calculated lifetime values

$$Y_i^{calc}(M_0, r_0, a_1, ..., a_m) = \left( \sum_{v''} A_{av''N}^{dv'N} \right)^{-1}, \qquad (11)$$

where the transition probabilities are expressed by the formula (4).

In the case of random experimental errors the least squares criterion (8) corresponds to the maximum likelihood principle. Obtaining the covariance matrix makes it possible to estimate uncertainties of semiempirical values obtained in the present work. These uncertainties include experimental errors, incompleteness of the limited set of experimental data and incomplete adequacy of used theoretical model to properties of a real molecule.

For calculations we used computer codes described in [10,13]. Details of such optimization-assisted solution of the inverse problem under the consideration will be described elsewhere. However, some important aspects have to be mentioned.

From the equations (9) and (10) it is visible that in the framework of our model the measured branching ratios may be used only for determination of the $r_0$, $m$, and $a_1$, $a_2$..., $a_m$ parameters of the $M_{da}(r)$ approximation. That gives us the *r*-dependence of the transition moment only in relative scale. At the same time, an optimal value of the $M_0$ parameter (determining absolute values of $M_{da}(r)$, Einstein coefficients (4) and lifetimes) should be obtained from experimental lifetime values. Therefore, an interactive procedure of minimizing the functional (8) was performed in two stages.

At the first stage, we used only experimental branching ratios listed in Table 2 as input data of the optimization procedure. Statistical Fisher and $\chi^2$ criteria as well as the cross-validation technique were employed. That gave to us an opportunity to find out that:

i) values of four branching ratios (marked by asterisks in Table 2) are in the contradiction with others in the framework of the theoretical model;

ii) m=2 is the optimal value of the parameter – a linear approximation is insufficient, while the third degree polynomial superfluous;

iii) $r$-dependencies of $M_{da}(r)$ obtained separately from the branching ratios of the H$_2$ and D$_2$ molecules coincide within error bars of their semiempirical determination, thus these two datasets from Table 2 may be used as a common input dataset for minimizing the functional (8);

iv) $r$-dependencies of $M_{da}(r)$ are practically independent from a value of $r_0$ in the range of internuclear distances in which overlapping of vibrational wave functions of upper and lower electronic-vibration states is sufficient, therefore $r_0$=0.9 Å was used in further calculations.

At the second stage, the electronic transition moment $M_{da}(r)$ (in an absolute scale) and corresponding covariance matrix were obtained by minimizing the functional (8) with the input dataset consisting of average values of the most reliable radiative lifetime values of the $d^3\Pi_u^-$, $v$=0-3, $N$ levels of the H$_2$ and branching ratios of the H$_2$ and D$_2$ molecules.

## 3. Conclusion

The optimization analysis briefly described above show applicability of adiabatic approximation and adequacy of our simplified theoretic model in reproducing the set of experimental data under the study. Therefore, we used adiabatic potential curves, the electronic transition moment $M_{da}(r)$, the model and the computer code reported in [10] for calculating Einstein coefficients for the $d^3\Pi_u^-$, $v'$, $N$ → $a^3\Sigma_g^+$, $v''$, $N$ spontaneous transitions and radiative lifetimes of upper rovibronic levels of the H$_2$, HD, D$_2$, and T$_2$ molecules. The results thus obtained are listed in Tables 3-8.

Currently available results of *ab initio* and less general non-adiabatic calculations are included into the tables and may be used for further analysis that is outside the scope of the present paper. However it is necessary to notice, that our semiempirical Einstein coefficients $A_{av''N}^{dv'N}$ are in good accordance with reported in [21] results of calculations based on the most precise and reliable *ab initio* calculations [22] of adiabatic potential curves and the electronic transition moment $M_{da}(r)$.

For the $d^3\Pi_u^-$, $v'$=4÷6, $N$=1 levels of the H$_2$ molecule the lifetimes caused by predissociation were estimated semiempirically for the first time.

## Acknowledgement

The present work was supported in part by Russian Foundation for Basic Research,
project No. 13-03-00786a.

# List of References

1. B. P. Lavrov, Opt. Spectrosc. **48**, 375 (1980).

2. A. I. Drachev, B. P. Lavrov, High. Temp. **26**, 129 (1988).

3. B. P. Lavrov, V. P. Prosikhin, Opt. Spectrosc. **58**, 317 (1985).

4. B. P. Lavrov, A. S. Melnikov, M. Kaening, J. Roepcke, Phys. Rev. E. **59**, 3526 (1999).

5. B. P. Lavrov, A. V. Pipa, J. Röpcke J. Plasma Sources Sci. Technol. **15**, 135 (2006).

6. B. P. Lavrov, A. S. Zhukov, Russ. J. Phys. Chem. B, **8**, 807 (2014).

7. *Nonthermal Plasma Chemistry and Physics*, Ed. by J. Meichsner, M. Schmidt, R. Schneider, H. E. Wagner (CRC Press Taylor & Francis Group, London, New York, Boca Raton, 2013), p. 206.

8. T. V. Kirbyateva, B. P. Lavrov, V. N. Ostrovsky, M. V. Tyutchev, V. I. Ustimov, Opt.Spectrosc. **52**, 21 (1982).

9. A. I. Drachev, B. P. Lavrov, V. P. Prosikhin, Vestn. Leningr. Univ. Fiz. & Khim. (USSR), Ser. 4 (No18), 16 (1988) ) [in Russian].

10. A. I. Drachev A.I., B. P. Lavrov, L. L. Pozdeev, V. P. Prosikhin, J. Struct. Chem. **30**, 337 (1989).

11. A. I. Drachev, B. P. Lavrov, V. P. Prosikhin, V. I. Ustimov, Sov. J. Chem. Phys. **4**, 1663 (1987).

12. B. P. Lavrov, L. L. Pozdeev, Opt.Spectrosc. **66,** 479 (1989).

13. A. I. Drachev, Lavrov B.P., L. L. Pozdeev, J.Struct.Chem. **31**, 482 (1990).

14. A. S. Melnikov, Ph.D. thesis, Leningrad State University, St. Petersburg, 1996 [in Russian].

15 *The Hydrogen Molecule Wavelength Tables of Gerhard Heinrich Dieke*, Ed. by H. M. Crosswhite (Wiley-Interscience, New York, London, Sydney, Toronto, 1972).

16. R. S. Freund, J. A. Schiavone, H. M. Crosswhite, J.Phys.Chem.Ref.Data. **14**, 235 (1985).

17. S. A. Astashkevich, B. P. Lavrov, A. V. Modin, I. S. Umrikhin, Russ. J. Phys. Chem. B, 2, 16 (2008).

18. K. P. Kureichik, A. I. Bezlepkin, A. S. Khomyak, V. V. Aleksandrov, *Gas Discharge Light Sources for Spectral Measurements* (Universitetskoe, Minsk, 1987) [in Russian].

19. V.S. Grebenkov, B. P. Lavrov, M.V. Tyutchev, Sov. J. Opt. Technol.**49**, 115 (1982).

20. T. Kiyoshima, H. Sato, S. O. Adamson, E. A. Pazyuk, A. V. Stolyarov, Phys.Rev.A **60**, 4494 (1999).

21. U. Fantz, V. Wuonderlich, Atomic Data and Nuclear Data Tables **92**, 853 (2006).

22. G. Staszewska, L. Wolniewicz, J. Mol. Spectrosc. **198**, 416 (1999)

23. P. Cahill, J. Opt. Soc. Am. **59**, 875 (1969).

24. M. A. Marechal, R. Jost, M. Lombardi, Phys. Rev.A **5**, 732 (1972).

25. R. S. Freund., F. A. Miller, J.Chem.Phys. **58**, 3565 (1973).

26. R. L. Day, R. J. Anderson, F. R. Sharpton, J.Chem.Phys. **69**, 5518 (1978).

27. I. P. Bogdanova, G. V. Efremova, B. P. Lavrov, V. N. Ostrovsky, V. I. Ustimov, V. I. Yakovleva, Opt.Spectrosc. **50**, 63 (1981).

28. A. I. Drachev, B. P. Lavrov, L. L. Pozdeev, Available from VINITI No. 6847-B87 (Moscow 1987), 46pp. [in Russian].

29. T. Kiyoshima, J. Phys. Soc. Jpn. **56**, 1989 (1987).

30. B. P. Lavrov, Dr. Science thesis, Leningrad State University, St. Petersburg, 1988 [in Russian].

31. J. A. Sanchez, J. Campos, J. Phys. (Paris) **49**, 445 (1988).

32. M. L. Burshtein, B. P. Lavrov, A. S. Melnikov, V. P. Prosikhin, S. V. Yurgenson, V. N. Yakovlev, Opt. Spectrosc. **68**, 166 (1990).

33. T. Kiyoshima, H. Sato, Phys.Rev. A **48**, 4771 (1993).

34. S. A. Astashkevich, B. P. Lavrov. J.Phys.Chem.Refer Data, **44**, 023105/1-42/ (2015).

35. B. P. Lavrov, V. P. Prosikhin, Opt. Spectrosc. **64**, 298 (1988).

Table 1. Optimal values of parameters for Hulbert-Hirschfelder potential curves of the $a^3\Sigma_g^+$ and $d^3\Pi_u^-$ electronic states of hydrogen molecule obtained in [14]. Standard deviations are shown in brackets in units of least significant digit.

| Electronic state | $U(r_e)$, cm$^{-1}$ | $D$, cm$^{-1}$ | $r_e$, $10^{-8}$ cm | $\beta$, $10^8$ cm$^{-1}$ | a | b |
|---|---|---|---|---|---|---|
| $a^3\Sigma_g^+$ | 93905.0(3) | 26723(30) | 0.98838(5) | 1.4111(7) | -0.373(6) | -0.192(1) |
| $d^3\Pi_u^-$ | 110698.4(4) | 23687(17) | 1.0498(2) | 1.3354(6) | -0.245(4) | -0.210(2) |

Table 2. Branching ratios $S_{av''}^{dv'}/S_{av'''}^{dv'}$ for Q-branch lines of Fulcher-α bands of $H_2$ and $D_2$ reported in various papers. Asterisks mark experimental data recognized as outliers during semiempirical analysis. Standard deviations are shown in brackets in units of least significant digit.

| $v'$ | $v''$ | $v'''$ | Experiment | | Semiempirical Determination | | *Ab initio* calculations | |
|---|---|---|---|---|---|---|---|---|
| | | | [12] | Present work | [12] | Present work | [20] | [21] |
| $H_2$ | | | | | | | | |
| 0 | 1 | 0 | 0,094 (5) | 0,109 (8) | 0,098 (8) | 0,106 (3) | 0,114 | 0,11262 |
| 1 | 0 | 1 | 0,056 (2) | 0,053 (2) | 0,054 (4) | 0,052 (2) | 0,047 | 0,04802 |
| 1 | 2 | 1 | 0,268 (15) | 0,256 (18) | 0,224 (18) | 0,242 (8) | 0,257 | 0,254 |
| 2 | 0 | 2 | 0,0038 (3) | 0,0034 (3) | 0,0036 (4) | 0,0033 (2) | 0,0027 | 0,00277 |
| 2 | 1 | 2 | 0,140 (5) | 0,130 (9) | 0,12 (1) | 0,118 (4) | 0,106 | 0,10773 |
| 2 | 3 | 2 | 0,41 (2) | 0,45 (3) | 0,39 (3) | 0,414 (13) | 0,439 | 0,43484 |
| 3 | 1 | 3 | 0,0168 (8) | ---- | 0,0135(13) | 0,0121(7) | 0,010 | 0,01029 |
| 3 | 2 | 3 | 0,219 (6) | 0,20 (1) | 0,21 (2) | 0,199 (6) | 0,179 | 0,18197 |
| 3 | 4 | 3 | 0,65 (3) | 0,71 (5) | 0,61 (5) | 0,64 (2) | 0,679 | 0,67191 |
| 4 | 3 | 4 | 0,340 (15) | 0,32 (2) | 0,32 (2) | 0,30 (1) | 0,270 | 0,27416 |
| 4 | 5 | 4 | 0,93 (6) | 1,00 (7) | 0,91 (7) | 0,95 (4) | 1,003 | 0,99005 |
| 5 | 4 | 5 | 0,44 (2) | 0,41 (3) | 0,47 (4) | 0,436 (16) | 0,382 | 0,38819 |
| 5 | 6 | 5 | 1,51 (8) | 1,64 (8)* | 1,32 (11) | 1,37 (7) | 1,444 | 1,4255 |
| 6 | 4 | 6 | 0,151 (9) | 0,135 (9) | 0,14 (1) | 0,120 (6) | 0,098 | 0,09998 |
| 6 | 5 | 6 | 0,62 (3) | 0,58 (4) | 0,65 (6) | 0,59 (3) | 0,519 | 0,52713 |
| $D_2$ | | | | | | | | |
| 0 | 1 | 0 | 0,145 (4) | 0,15(1)* | 0,13 (1) | 0,134 (4) | 0,143 | 0,1419 |
| 1 | 0 | 1 | 0,084 (5) | 0,084 (5) | 0,090 (7) | 0,092 (3) | 0,0825 | 0,08359 |
| 1 | 2 | 1 | 0,310 (9) | 0,33 (2)* | 0,30 (2) | 0,31 (1) | 0,335 | 0,3323 |
| 2 | 1 | 2 | 0,210 (6) | 0,20 (1) | 0,21 (2) | 0,21 (1) | 0,187 | 0,18965 |
| 2 | 3 | 2 | 0,57 (3) | 0,61 (3) | 0,55 (4) | 0,57(1) | 0,599 | 0,59501 |
| 3 | 2 | 3 | 0,38 (1) | 0,36 (2) | 0,36 (3) | 0,355 (7) | 0,324 | 0,32814 |
| 3 | 4 | 3 | 0,91 (5) | 0,97 (5) | 0,90 (7) | 0,95 (2) | 0,977 | 0,97041 |
| 4 | 3 | 4 | 0,56 (3) | 0,54 (3) | 0,58 (4) | 0,56 (1) | 0,507 | 0,51489 |
| 4 | 5 | 4 | 1,62 (15) | 1,7 (1)* | 1,45 (12) | 1,51 (3) | 1,539 | 1,53037 |
| 5 | 4 | 5 | 0,85 (4) | 0,81 (4) | 0,90 (7) | 0,86 (1) | 0,767 | 0,7782 |
| 5 | 6 | 5 | 2,49 (15) | 2,6 (2) | 2,32 (19) | 2,41 (1) | 2,422 | 2,40789 |

Table 3. The lifetimes $\tau_{dvN}$ of the $d^3\Pi_u^-, v, N$ electronic-vibro-rotational levels of the H$_2$ molecule reported in various papers. Most reliable experimental data are printed in bold face; their averages (Expt) are italicized.

| N | \multicolumn{7}{c}{$\tau_{dvN}$, ns} | Method | Ref. |
|---|---|---|---|---|---|---|---|---|---|
|   | $v=0$ | 1 | 2 | 3 | 4 | 5 | $v=6$ | | |
| \multicolumn{10}{c}{H$_2$} |
| 1-3 | 68±5 | 58±5 | 62±5 | - | - | - | - | PS | [23] |
| 2 | \multicolumn{4}{c|}{31.5±3} | - | - | - | MH | [24] |
| 1 | \multicolumn{4}{c|}{32±5} | - | - | - | OMR | [25] |
| 1-3 | 42.2±2.6 | 39.0±2.3 | 39.0±2.3 | 39.6±2.6 | - | - | - | DC | [26] |
| 1 | **39.4±2.0** | **36.5±1.9** | **36.2±2.0** | **36.4±2.1** | - | - | - | | |
| 2 | 44.2±2.7 | 42.0±2.5 | 40.5±2.5 | 43.7±3.2 | - | - | - | | |
| 3 | 43.3±3.0 | **38.8±2.6** | **40.6±2.5** | **39.5±2.5** | - | - | - | | |
| 1,3 | \multicolumn{4}{c|}{35±5} | - | - | - | DC | [27] |
| 1 | 39.6±2.0 | 39.6±2.0 | 40.0±2.0 | - | - | - | - | SDRL | [8] |
| 1 | 39.3±3.2 | 39.8±3.2 | 40.4±3.2 | 41.1±3.4 | 41.8±3.6 | - | - | SDRL | [9] |
| 1 | 38.7±2.1 | 39.7±2.1 | 40.9±2.2 | 42.2±2.2 | - | - | - | SDRL | [28] |
| 1 | **40.8±2.0** | - | - | - | - | - | - | DC | [29] |
| 2 | 41.8±3.0 | **40.7±1.9** | **39.0±2.9** | **40.2±3.4** | - | - | - | | |
| 1 | - | - | - | - | 16±2 | | | plasma | [30] |
| 1 | 30.0±2.1 | 26.8±1.9 | 22.3±1.6 | - | - | - | - | DC | [31] |
| 1,3 | **39.5±1.2** | **38.2±0.8** | **42.2±0.8** | **40±3** | **19±1** | **15±1** | **16±3** | DC | [32] |
| 1 | **39.1±1.4** | **37.7±1.3** | **37.8±1.2** | **38.3±1.3** | - | - | - | DC | [33] |
| 1,3 | - | - | - | - | **16.3±0.7** | **11±1** | **11±3** | DC | [20] |
| 1 | 32.8 | 33.9 | 34.2 | 34.5 | 34.9 | 35.2 | 35.6 | AA | |
| 1,3 | 37.8 | 38.2 | 38.8 | 39.1 | 10.2±0.5 | 7.5±0.3 | 5.8±0.3 | NAM | |
| 0 | 38.848 | 39.071 | 39.140 | 39.704 | 40.000 | 39.805 | 40.168 | AA | [21] |
| 1,3 | *39.7±0.5* | *38.6±0.9* | *39.4±1.5* | *39.1±1.3* | *17.6±1.5* | *13±2* | *14±3* | Expt | Present work |
| 1 | 37.9±0.5 | 38.7±0.5 | 39.5±0.5 | 40.5±0.6 | 41.5±0.6 | 42.7±0.8 | 44±1 | SDRL | |
| 1 | - | - | - | - | (31±6) | (19±5) | (22±8) | Diss | |

Table 4. The lifetimes $\tau_{dvN}$ of the $d^3\Pi_u^-, v, N$ electronic-vibro-rotational levels of the $D_2$, HD and $T_2$ molecules reported in various papers.

| N | $\tau_{dvN}$, ns | | | | | | | Method | Ref. |
|---|---|---|---|---|---|---|---|---|---|
|   | v=0 | 1 | 2 | 3 | 4 | 5 | v=6 |   |   |
| $D_2$ | | | | | | | | | |
| 1 | 39.4±3.2 | 39.8±3.1 | 40.2±3.2 | 40.6±3.3 | - | - | - | SDRL | [9] |
| 1 | 38.5±2.0 | 39.2±2.1 | 40.0±2.1 | 40.8±2.1 | 41.7±2.2 | - | - | SDRL | [28] |
| 1 | - | 38.4±1.5 | - | - | - | - | - | DC | [33] |
| 2 | - | 39.2±1.4 | 40.1±1.5 | - | - | - | - | | |
| 3 | - | 39.7±1.8 | - | - | - | - | - | | |
| 1 | - | - | - | - | 37.4±0.6 | 36±3 | 37±3 | DC | [20] |
| 1 | 33.8 | 34.0 | 34.2 | 34.4 | 34.5 | 34.7 | 34.9 | NAM | |
| 0 | 38.799 | 39.021 | 39.216 | 39.556 | 39.730 | 39.831 | 40.253 | AA | [21] |
| 1 | 36.6±0.5 | 37.1±0.5 | 37.7±0.4 | 38.2±0.4 | 38.9±0.3 | 39.5±0.3 | 40.2±0.4 | SDRL | Present work |
| HD | | | | | | | | | |
| 1 | 39.4±3.2 | 39.8±3.2 | 40.3±3.2 | 41.2±3.3 | - | - | - | SDRL | [9] |
| 1 | 38.6±2.0 | 39.5±2.1 | 40.5±2.1 | 41.5±2.2 | 42.7±2.3 | - | - | SDRL | [28] |
| 0 | 38.826 | 39.059 | 39.211 | 39.669 | 39.860 | 39.866 | 40.240 | AA | [21] |
| 1 | 36.7±0.5 | 37.3±0.4 | 38.0±0.4 | 38.8±0.3 | 39.6±0.3 | 40.5±0.4 | 41.4±0.6 | SDRL | Present work |
| $T_2$ | | | | | | | | | |
| 0 | 38.776 | 38.974 | 39.167 | 39.425 | 39.596 | 39.760 | 40.060 | AA | [21] |
| 1 | 36.6±0.5 | 37.8±0.5 | 38.2±0.4 | 38.7±0.4 | 39.1±0.4 | 39.1±0.3 | 40.1±0.3 | SDRL | Present work |

Table 5. Einstein coefficients for the $d^3\Pi_u^-, v',N=1 \rightarrow a^3\Sigma_g^+, v'',N=1$ electronic-vibro-rotational spontaneous transitions of the H$_2$ molecule, obtained semiempirically in the present work and results of non-empirical adiabatic calculations (in italics) reported in [21]. The uncertainties of the semiempirical determination (one standard deviation) are shown in brackets in units of least significant digit.

| | $A_{av''1}^{dv'1}$, $10^7$ s$^{-1}$ | | | | | | |
|---|---|---|---|---|---|---|---|
| v″ | v′=0 | v′=1 | v′=2 | v′=3 | v′=4 | v′=5 | v′=6 |
| 0 | 2.48 (3) *2.4077* | 0.168 (4) *0.15258* | 0.0128 (7) *0.010712* | 0.00112 (14) *0.00083952* | 0.00010 (3) *0.000058695* | 0.000011 (6) *0.0000038977* | 0.000 *3.5846E-9* |
| 1 | 0.161 (3) *0.16552* | 2.10 (3) *2.0655* | 0.311 (6) *0.28369* | 0.0372 (16) *0.031899* | 0.0046 (5) *0.0036443* | 0.00059 (11) *0.00039449* | 0.00007 (3) *0.000045831* |
| 2 | 0.00077 (4) *0.00092743* | 0.316 (6) *0.32649* | 1.74 (2) *1.7377* | 0.422 (7) *0.38874* | 0.071 (3) *0.062233* | 0.0115 (9) *0.0095638* | 0.0019 (3) *0.0014215* |
| 3 | 0.0000 (0) *7.7501E-9* | 0.00249 (13) *0.0029732* | 0.458 (8) *0.47993* | 1.42 (2) *1.4317* | 0.499 (8) *0.46420* | 0.112 (3) *0.098902* | 0.0228 (13) *0.019437* |
| 4 | 0.0000 (0) *5.6159E-9* | 0.0000 (0) *2.8248E-7* | 0.0052 (4) *0.00623109* | 0.59 (1) *0.62363* | 1.12 (2) *1.1529* | 0.541 (8) *0.50844* | 0.154 (3) *0.13818* |
| 5 | 0.0000 (0) *2.1645E-12* | 0.0000 (0) *1.8340E-8* | 0.0000 (0) *2.1093E-6* | 0.0088 (7) *0.010633* | 0.701 (14) *0.75563* | 0.86 (2) *0.90656* | 0.548 (8) *0.52167* |
| 6 | 0.0000 (0) *1.2126E-11* | 0.0000 (0) *4.0941E-13* | 0.0000 (0) *6.3848E-8* | 0.0000 (0) *1.0579E-5* | 0.0127 (13) *0.015873* | 0.80 (2) *0.87420* | 0.65 (2) *0.69697* |
| 7 | 0.0000 (0) *2.4635E-11* | 0.0000 (0) *4.9220E-10* | 0.0000 (0) *1.1783E-9* | 0.0000 (0) *1.8096E-7* | 0.00005 (1) *0.000043242* | 0.016 (2) *0.021225* | 0.88 (3) *0.97831* |

Table 6. Einstein coefficients for the $d^3\Pi_u^-, v', N=1 \rightarrow a^3\Sigma_g^+, v'', N=1$ electronic-vibro-rotational spontaneous transitions of the HD molecule, obtained semiempirically in the present work and results of non-empirical adiabatic calculations (in italics) reported in [21]. The uncertainties of the semiempirical determination (one standard deviation) are shown in brackets in units of least significant digit.

| | $A_{av''1}^{dv'1}$, $10^7$ s$^{-1}$ | | | | | | |
|---|---|---|---|---|---|---|---|
| v″ | v′=0 | v′=1 | v′=2 | v′=3 | v′=4 | v′=5 | v′=6 |
| 0 | 2.53 (3) *2.3816* | 0.200 (4) *0.17675* | 0.0166 (8) *0.013606* | 0.00159 (15) *0.0011678* | 0.00018 (3) *0.000094558* | 0.000021 (7) *0.0000097204* | 0.0000 (0) *3.4221E-7* |
| 1 | 0.195 (3) *0.19218* | 2.096 (17) *1.9963* | 0.362 (5) *0.32177* | 0.0470 (17) *0.039228* | 0.0062 (5) *0.0047681* | 0.00084 (12) *0.00056369* | 0.00011 (3) *0.000077136* |
| 2 | 0.00164 (4) *0.0017989* | 0.377 (6) *0.37403* | 1.699 (11) *1.6381* | 0.482 (6) *0.43271* | 0.088 (3) *0.074457* | 0.015 (1) *0.011897* | 0.0026 (3) *0.0018632* |
| 3 | 0.0000 (0) *3.5972E-7* | 0.00511 (15) *0.0056742* | 0.543 (7) *0.54255* | 1.343 (7) *1.3127* | 0.561 (7) *0.50823* | 0.134 (3) *0.11556* | 0.0283 (14) *0.023311* |
| 4 | 0.0000 (0) *1.1262 E-8* | 0.0000 (0) *9.6306 E-7* | 0.011 (4) *0.011778* | 0.688 (7) *0.69586* | 1.030 (7) *1.0237* | 0.600 (7) *0.54897* | 0.182 (3) *0.15857* |
| 5 | 0.0000 (0) *1.2125 E-9* | 0.0000 (0) *3.7675 E-8* | 0.0000 (0) *1.4824 E-6* | 0.0184 (8) *0.020115* | 0.814 (9) *0.83245* | 0.766 (9) *0.77428* | 0.602 (7) *0.55742* |
| 6 | 0.0000 (0) *3.0850 E-10* | 0.0000 (0) *2.7139 E-10* | 0.0000 (0) *1.5520 E-7* | 0.0000 (0) *9.9268 E-7* | 0.0276 (15) *0.030502* | 0.918 (12) *0.95108* | 0.55 (1) *0.56619* |
| 7 | 0.0000 (0) *4.8484 E-10* | 0.0000 (0) *8.3861 E-10* | 0.0000 (0) *1.1240 E-9* | 0.0000 (0) *4.9952 E-7* | 0.0000 (0) *3.3328 E-8* | 0.038 (3) *0.042424* | 1.00 (2) *1.0518* |

Table 7. Einstein coefficients for the $d^3\Pi_u^-, v', N=1 \rightarrow a^3\Sigma_g^+, v'', N=1$ electronic-vibro-rotational spontaneous transitions of the D$_2$ molecule, obtained semiempirically in the present work and results of non-empirical adiabatic calculations (in italics) reported in [21]. The uncertainties of the semiempirical determination (one standard deviation) are shown in brackets in units of least significant digit.

| | $A_{av''1}^{dv'1}$, $10^7$ s$^{-1}$ | | | | | | |
|---|---|---|---|---|---|---|---|
| v'' | v'=0 | v'=1 | v'=2 | v'=3 | v'=4 | v'=5 | v'=6 |
| 0 | 2.48 (3) *2.3387* | 0.242 (4) *0.21551* | 0.0227 (8) *0.018763* | 0.00233(16) *0.0017678* | 0.00026 (3) *0.00016204* | 0.000032 (6) *0.000020439* | 0.0000 (0) *1.6819E-6* |
| 1 | 0.242 (4) *0.23479* | 1.98 (2) *1.8841* | 0.425 (6) *0.38098* | 0.0619(18) *0.052206* | 0.0087 (5) *0.0068073* | 0.00126 (13) *0.00088756* | 0.00019 (3) *0.00012672* |
| 2 | 0.00375 (5) *0.0039083* | 0.458 (6) *0.4473* | 1.539 (11) *1.4795* | 0.552 (7) *0.49835* | 0.112 (3) *0.095818* | 0.020 (1) *0.01626* | 0.0036 (3) *0.0026485* |
| 3 | 0.0000 (0) *6.9282E-6* | 0.01162 (18) *0.012047* | 0.646 (8) *0.63500* | 1.156 (7) *1.1276* | 0.626 (7) *0.57000* | 0.1670 (4) *0.14449* | 0.0370 (15) *0.030516* |
| 4 | 0.0000 (0) *6.7372E-9* | 0.000026 (4) *0.000027294* | 0.0237 (4) *0.024539* | 0.803 (9) *0.79698* | 0.836 (5) *0.82852* | 0.6543 (7) *0.60063* | 0.221 (4) *0.19329* |
| 5 | 0.0000 (0) *1.9986E-9* | 0.0000 (0) *4.3313E-8* | 0.000052 (8) *0.000069366* | 0.0398 (9) *0.041361* | 0.931 (9) *0.93260* | 0.5759 (5) *0.58234* | 0.643 (6) *0.5959* |
| 6 | 0.0000 (0) *2.9027E-11* | 0.0000 (0) *3.4305E-9* | 0.0000 (0) *2.0363E-7* | 0.00011 (2) *0.00013613* | 0.0597 (16) *0.062381* | 1.029 (1) *1.0416* | 0.374 (6) *0.38674* |
| 7 | 0.0000 (0) *7.7599E-10* | 0.0000 (0) *3.5382E-9* | 0.0000 (0) *2.0664E-10* | 0.0000 (0) *6.0937E-7* | 0.00016 (4) *0.00022215* | 0.0826 (3) *0.087127* | 1.100 (11) *1.1257* |

Table 8. Einstein coefficients for the $d^3\Pi_u^-, v', N=1 \rightarrow a^3\Sigma_g^+, v'', N=1$ electronic-vibro-rotational spontaneous transitions of the T$_2$ molecule, obtained semiempirically in the present work and results of non-empirical adiabatic calculations (in italics) reported in [21]. The uncertainties of the semiempirical determination (one standard deviation) are shown in brackets in units of least significant digit.

| | $A_{av''1}^{dv'1}$, $10^7$ s$^{-1}$ | | | | | | |
|---|---|---|---|---|---|---|---|
| v″ | v′ =0 | v′ =1 | v′ =2 | v′ =3 | v′ =4 | v′ =5 | v′ =6 |
| 0 | 2.43 (3) *2.2875* | 0.291 (4) *0.26025* | 0.0304 (8) *0.025648* | 0.00336(17) *0.0025841* | 0.00039 (3) *0.00027192* | 0.000059 (7) *0.000030817* | 0.0000 (0) *3.7815×10$^{-6}$* |
| 1 | 0.296 (4) *0.28385* | 1.846 (16) *1.7536* | 0.496 (6) *0.44614* | 0.0811(19) *0.069105* | 0.0122 (5) *0.0096843* | 0.00183 (13) *0.0013283* | 0.00029 (3) *0.00019768* |
| 2 | 0.00746 (9) *0.0075201* | 0.547 (7) *0.52730* | 1.35 (1) *1.3010* | 0.623 (7) *0.56584* | 0.143 (3) *0.12314* | 0.027 (1) *0.022297* | 0.0050 (3) *0.0037825* |
| 3 | 0.000036 (2) *0.000039579* | 0.0225 (3) *0.022655* | 0.751 (9) *0.72945* | 0.950 (6) *0.92680* | 0.686 (7) *0.62735* | 0.207 (4) *0.18060* | 0.0488 (16) *0.040614* |
| 4 | 0.0000 (0) *3.0605E-9* | 0.000145 (7) *0.0001624* | 0.0450 (5) *0.04515* | 0.91 (1) *0.89175* | 0.629 (3) *0.62635* | 0.694 (7) *0.64071* | 0.268 (4) *0.23544* |
| 5 | 0.0000 (0) *1.4405E-8* | 0.0000 (0) *1.8895E-8* | 0.00039 (2) *0.00042155* | 0.0742(9) *0.074628* | 1.03 (1) *1.0151* | 0.387 (3) *0.39475* | 0.660 (6) *0.61524* |
| 6 | 0.0000 (0) *2.4326E-9* | 0.0000 (0) *2.7242E-9* | 0.0000 (0) *1.7555E-8* | 0.00085 (5) *0.00085796* | 0.114 (2) *0.11066* | 1.14 (1) *1.1020* | 0.221 (3) *0.22459* |
| 7 | 0.0000 (0) *5.2151E-9* | 0.0000 (0) *1.7385E-9* | 0.0000 (0) *6.5876 E-9* | 0.0000 (0) *3.3341E-8* | 0.0013 (1) *0.0015178* | 0.150 (3) *0.15223* | 1.14 (1) *1.1557* |
| 8 | 0.0000 (0) *3.8472E-9* | 0.0000 (0) *1.4492E-9* | 0.0000 (0) *1.3189E-10* | 0.0000 (0) *2.2571E-8* | 0.0000 (0) *6.3543 E-10* | 0.0021 (2) *0.0024305* | 0.194 (3) *0.19871* |